\documentclass[twocolumn,superscriptaddress,showpacs,oneside,amsmath,amssymb,prb]{revtex4}
\usepackage{epsfig,amssymb,amsfonts}
\def\Re{\mathop{\rm Re}}
\def\Im{\mathop{\rm Im}}
\begin{document}

\title{ Band structure renormalization and weak pseudogap behavior in
  Na$_{0.33}$CoO$_2$: Fluctuation exchange study based on a single
  band model}

\author{Zi-Jian Yao}
\affiliation{National Laboratory of Solid State Microstructures and
  Department of Physics, Nanjing University, Nanjing 210093, China}
\affiliation{Department of Physics and Center of Theoretical and
  Computational Physics, The University of Hong Kong, Pokfulam Road,
  Hong Kong, China}
\author{Jian-Xin Li}
\affiliation{National Laboratory of Solid State Microstructures and
  Department of Physics, Nanjing University, Nanjing 210093, China}
\author{Z. D. Wang}
\affiliation{Department of Physics and Center of Theoretical and
  Computational Physics, The University of Hong Kong, Pokfulam Road,
  Hong Kong, China} \affiliation{National Laboratory of Solid State
  Microstructures and Department of Physics, Nanjing University,
  Nanjing 210093, China}

\date{\today}
\begin{abstract}
  Based on a single band Hubbard model and the fluctuation exchange
  approximation, the effective mass and the energy band
  renormalization in Na$_{0.33}$CoO$_2$ is elaborated. The
  renormalization is observed to exhibit certain kind of anisotropy,
  which agrees qualitatively with the angle-resolved photoemission
  spectroscopy (ARPES) measurements. Moreover, the spectral function
  and density of states (DOS) in the normal state are calculated, with
  a weak pseudogap behavior being seen, which is explained as a result
  of the strong Coulomb correlations. Our results suggest that the
  large Fermi surface (FS) associated with the $a_{1g}$ band plays
  likely a central role in the charge dynamics.
\end{abstract}


\maketitle The layered oxide Na$_x$CoO$_2$ has attracted much
attention due to its possible connection to the high-$T_c$
superconductivity since the discovery of superconductivity in
it.~\cite{Takada_03,Wang_03} This material consists of two-dimensional
CoO$_{2}$ layers, where Co's form a triangular lattice, making the
Na$_x$CoO$_2$ a possible realization of Anderson's resonating valence
bond (RVB) state.~\cite{Baskaran_03} By hydration, it becomes a
superconductor with $T_c\approx 5$K in Na$_{0.337}$CoO$_2
\cdot$1.3H$_{2}$O~\cite{Takada_03}.
The unhydrated compound Na$_x$CoO$_2$ exhibits a rich phase diagram: a
paramagnetic metal ($x<0.5$), a charge-ordered insulator ($x=0.5$), a
"Curie-Weiss metal" ($x\sim0.7$), and a magnetic order state
($x\geqq0.75$).~\cite{Foo_04}


Experimentally, it is indicated that Na$_x$CoO$_2$ seems to be
strongly electronic correlated.~\cite{Wang_03,Jin_03,Chou_04}
Angle-resolved photoemission spectroscopy (ARPES) measurements
show a strong mass renormalization (The effective mass is about
5-10 times larger than the bare mass.) for $x=0.6$.~\cite{Yang_04}
Decreasing the Na concentration to $x=0.3$, the cobaltates appear
to be less electron-correlated with a weaker but still apparent
energy band renormalization factor ($\sim2$).~\cite{Yang_05} For
the hydrated compounds, the effective mass is estimated to be
$\sim2-3$.~\cite{Shimojima_06} In addition, recent experiments
reported that Na$_{0.33}$CoO$_2\cdot$1.3H$_2$O display certain
pseudogap behaviors such as the decreasing of the Knight
shift~\cite{Ning_04} and the density of states (DOS) at the Fermi
level below 300K.~\cite{Shimojima_05} Optical spectroscopy
measurements for Na$_{0.2}$CoO$_2$ and Na$_{0.5}$CoO$_2$ also
suggest the incipient formation of pseudogap.~\cite{Hwang_05}
However, unlike the pseudogap effect observed in the high-$T_c$
cuprates, the pseudogap behavior is rather weak. The
renormalization of the energy band and the pseudogap formation are
directly related to the electronic structure, such as the
quasiparticle spectral function, the quasiparticle dispersion, and
the Fermi surface (FS) topology. Therefore it is of importance and
significance to study the normal state quasiparticle dynamics.

It is believed that the topology of the FS plays an important role in
the unconventional superconductivity. The local density approximation
(LDA) calculations~\cite{Singh_00,Lee_04} predict a large FS
associated with the $a_{1g}$ band and six pockets associated with the
$e'_g$ band. However, intriguingly, the small pockets have not been
observed in the ARPES measurements.~\cite{Yang_05, Hasan_04,
  Shimojima_06} The unexpected inconsistency of the topology of the FS
between the LDA results and the ARPES measurements has aroused much
controversy. One possibility might be due to the surface effect in the
ARPES measurements. From another viewpoint, Zhou $\it et$ $al.$
suggested that the strong Coulomb interactions may induce significant
renormalization of the band structure which pulls the $e'_g$ band down
from the Fermi energy.~\cite{Zhou_05} Furthermore, the $e'_g$ band is
suggested to be relevant to the spin-triplet superconductivity in some
theoretical and numerical studies.~\cite{Kuroki_04,Kuroki_05} On the
other hand, the possibility of the spin-singlet superconductivity is
also suggested according to recent Knight-shift
measurements.~\cite{Kobayashi_03,Zheng_06} Taking the above
considerations into account, 
a single band model focusing on the $a_{1g}$ band may be a starting
point as a minimal model of Na$_x$CoO$_2$.

In this paper, we study the normal state electronic structure in
Na$_{0.33}$CoO$_2$ on the basis of the single band Hubbard model
and the fluctuation exchange (FLEX) approximation. Several
experimental features, such as the band renormalization and the
weak pseudogap behavior, are well reproduced, which suggests an
important role of the holelike Fermi surface centered around the
$\Gamma$ point in the quasiparticle dynamics. We start with the
two-dimensional single band Hubbard model given by,
\begin{equation}
  H=\sum_{ij,\sigma}(t_{ij}c_{i\sigma}^{\dagger}c_{j\sigma}+h.c.)+U\sum_{i}n_{i\uparrow}n_{i\downarrow}-\mu\sum_{i\sigma}n_{i\sigma}
\end{equation}
where $t_{ij}$ denotes the hopping term (in the following we will
use $t_1,t_2,t_3$ and $t_4$ to denote the hoppings along the
nearest- to the fourth nearest-neighbors), $U$ the on-site Coulomb
repulsion and $\mu$ the chemical potential. To reproduce the large
FS around the $\Gamma$ point, we set the parameters ($t_1$, $t_2$,
$t_3$, $t_4$) of the bare dispersion $\epsilon_{\mathbf{k}}$ to be
(-1, 0, 0, 0.2), where $t_1$ is set to be the unit hereafter.
According to the numerical result for the
bandwidth,~\cite{Singh_00,Kunes_03} one can get $t_1=130$ meV. We
note that the main feature of the bare dispersion is similar to that
of the band with an $a_{1g}$ character in the LDA calculation by Lee
{\it et al.}.~\cite{Lee_04}

\begin{figure}
  \epsfig{figure=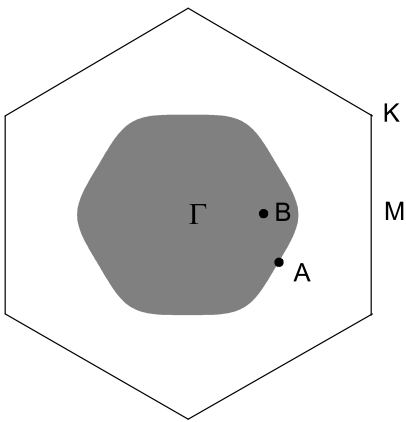,width=0.3\textwidth}
  \caption{The Fermi surface of Na$_{0.33}$CoO$_2$ for
    ($t_1$,$t_2$,$t_3$ $t_4$)=(-1,0,0,0.2). The grey area denotes the
    unoccupied electron states. Point A near the Fermi surface
    indicates the $\mathbf{k}$-point where $m^*$ is calculated.}
\end{figure}

The FLEX approximation is employed in our study, which has been
applied to the two dimensional Hubbard model in
literatures.~\cite{Monthoux_94,Pao_94,Dahm_95,Yao_06} As a
self-consistent approximation, the FLEX approximation solves the
Dyson's equation with a primarily RPA+ladder type effective
interaction self-consistently. Based on the scenario of the FLEX
approximation, the self-energy is given by,
\begin{equation}\label{sigma}
  \Sigma(k)=\frac{T}{N}\sum_{q}V_{eff}(k-q)G(q),
\end{equation}
where
\begin{equation}\label{Veff}
  V_{eff}(q)=\frac{3}{2}U^2\chi_s(q)+\frac{1}{2}U^2\chi_c(q).
\end{equation}
The spin susceptibility is
\begin{equation}
  \chi_s({q})=\frac{\overline{\chi}(q)}{1-U\overline{\chi}(q)},
\end{equation}
and the charge susceptibility is
\begin{equation}\label{chic}
  \chi_c({q})=\frac{\overline{\chi}(q)}{1+U\overline{\chi}(q)},
\end{equation}
with the irreducible susceptibility,
\begin{equation}
\overline{\chi}(q)=-\frac{T}{N}\sum_{k}G(k+q)G(k).
\end{equation}
The electron Green's function is give by,
\begin{equation}\label{dyson}
  G(k)=[i\omega_n-\epsilon_{k}-\Sigma(k)]^{-1}.
\end{equation}
In the above equations, $k\equiv(\mathbf{k},i\mathbf{\omega}_n)$ and
$q\equiv(\mathbf{q},i\mathbf{\nu}_n)$ are used, $T$ is the
temperature. These equations are solved self-consistently, where
$N=64\times64$ $\mathbf{k}$-point meshes and up to 2048 Matsubara
frequencies $\omega_n=(2n+1){\pi}T$ are taken. The electron density is
determined by the chemical potential $\mu$ from the equation
$<n>=1-\frac{2T}{N}\sum_kG(k)$.

\begin{figure}
  \epsfig{figure=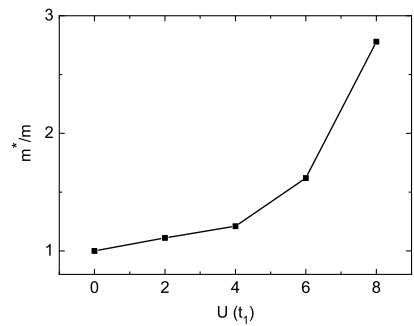,width=0.3\textwidth}
  \caption{The ratio of the effective mass to the bare mass $m^*/m$ at
    point A as indicated in Fig. 1 versus the on-site Coulomb
    repulsion $U$ at $T=0.1$.}
\end{figure}

As an important physical quantity related closely to the strength of
the electronic correlations, the effective mass $m^*$ is given by
\begin{equation}
  \frac{m}{m^*}=Z(1+\frac{m}{k_F}\frac{\partial}{\partial{k}}\Re\Sigma(k,0)\Big|_{k=k_F}),
\end{equation}
where
\begin{equation}
  Z=(1-\frac{\partial}{\partial\omega}\Re\Sigma(k_F,\omega)\Big|_{\omega=0})^{-1}
\end{equation}
is the renormalization constant. The self-energy with real frequencies
$\Sigma(\mathbf{k},\omega)$ is obtained with the Pad\'{e}
approximants,~\cite{Vidberg_77} which analytically continue
$\Sigma(\mathbf{k},i\omega)$ from the Matsubara frequencies to the
real-frequency axis. The ratio of the effective mass to the bare mass
$m^*/m$ with different on-site Coulomb repulsions $U$ near the FS
(point A in Fig. 1) is plotted in Fig. 2, where $m^*/m$ increases
monotonously with the increasing of $U$. It is noticeable that when
$U$ is less than 4, the enhancement of the effective mass is
inapparent ($<1.2$). With a strong Coulomb repulsion ($U\geq6\approx
0.8$ eV), the effective mass is enhanced significantly. Comparing the
calculated effective mass with the ARPES measurements ($\sim2$ for
$x\sim0.3$), we set $U\sim6$ in the present simple model to reflect
the strong correlation effect in Na$_{0.33}$CoO$_2$.~\cite{U} The
strong electronic correlations are evident from Fig. 3 (a), where the
sharp quasiparticle peak is suppressed with strong Coulomb repulsion.
For $U=2$ and $T=0.1$, the spectral function shows a sharp peak near
the Fermi energy. Increasing the Coulomb repulsion to $U=6$, the
spectral function is broadened and the spectral weight at the Fermi
level is reduced greatly.

\begin{figure}
  \epsfig{figure=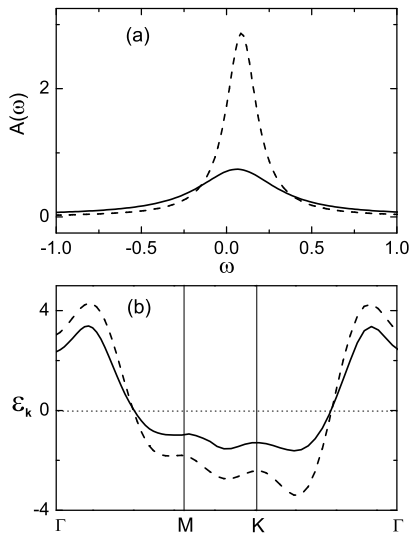,width=0.3\textwidth}
  \caption{ (a) The effect of the on-site Coulomb repulsion on the
    spectral function at point A of Fig. 1. Solid line: $U=6$, $T=0.1$.
    Dashed line: $U=2$, $T=0.1$. (b) Solid line: the renormalized band
    dispersion for $U=6$ and $T=0.1$. Dashed line: the bare band
    dispersion. Dotted line: the Fermi energy.}
\end{figure}

For $U=6$ 
we now investigate the renormalization of the quasiparticle energy
band. To do it, we need to evaluate the spectral function of
quasiparticles defined as
\begin{equation}
  A(\mathbf{k},\omega)=-\frac{1}{\pi}\Im{G(\mathbf{k},\omega)},
\end{equation}
where $G(\mathbf{k},\omega)$ is the dressed Green's function in real
frequecy, which is analytically continued from the dressed Green's
function in imaginary-frequecy via the Pad\'{e} approximants. Then,
the renormalized band dispersion is determined by the position of
the peak of $A(\mathbf{k},\omega)$. The calculated result at $T=0.1$
is shown in Fig. 3 (b), where the solid and the dotted lines denote
the renormalized and the bare energy bands, respectively. One can
see that the bandwidth is compressed obviously due to the strong
Coulomb interactions. From Fig. 3 (b) we get the bandwidth of the
bare band dispersion ($7.7$) and the renormalized bandwidth
($5.0$), which gives a band renormalization factor $1.54$.
This is consistent with the ARPES measurements for
$x=0.3$.~\cite{Yang_05} Moreover, we find that the band
renormalization shows an anisotropy along the $\Gamma$-$K$ and
$\Gamma$-$M$ direction, namely the renormalization along the
$\Gamma$-$K$ direction is stronger than that along the $\Gamma$-$M$
direction. The similar anisotropy has been found in the ARPES
experiments for $x=0.6$,~\cite{Yang_04} though the experimental
result seems stronger than what we disclosed here. We note that the
compound Na$_x$CoO$_2$ with $x=0.6$ is in fact in the range of the
Curie-Weiss metal, in which the correlation is stronger than that in
the paramagnetic metal with $x=0.3$. So, a weaker anisotropy in the
renormalization is expected. This anisotropy originates from the
nesting of the Fermi surface along the $\Gamma$-$K$ direction as
shown in Fig. 1.

\begin{figure}
  \epsfig{figure=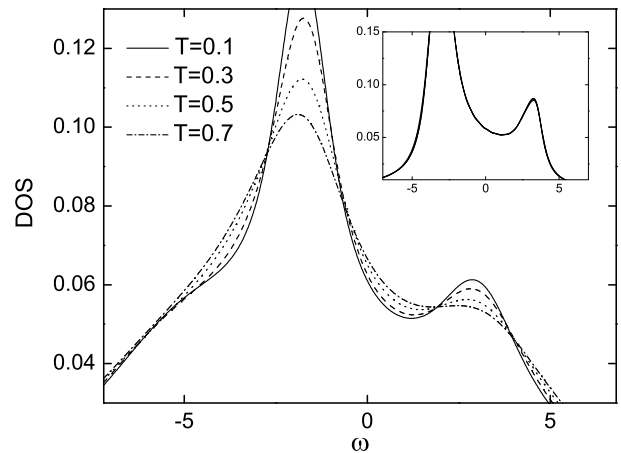,width=0.45\textwidth}
  \caption{ The total density of states (DOS) versus energy for $U=6$.
    Solid line: $T=0.1$, dashed line: $T=0.3$, dotted line: $T=0.5$, and dash-dotted line:
    $T=0.7$. Inset: DOS plots for $U=2$ at $T=0.1, 0.3, 0.5, 0.7$.}
\end{figure}

We now turn to address the weak pseudogap behavior. This is manifested
in the suppression of the density of states at the Fermi level. We
present the $\omega$ dependence of the density of states at different
temperatures in Fig. 4. With the decrease of temperature from $T=0.7$
(dash dotted) through $T=0.5$ (dotted) and $T=0.3$ (dashed) to $T=0.1$
(solid), a weak suppression of the density of states near the Fermi
energy is evident, suggesting the opening of a weak pseudogap. The
weak pseudogap behavior is also manifested in the quasiparticle
spectral function $A(\mathbf{k},\omega)$ which measures the
probability to find the quasiparticle at momentum $\mathbf{k}$ and
frequency $\omega$. As shown in Fig. 5 (a), with the decrease of
temperature (from the dashed line to the solid line), the spectral
weight is transferred away from the region around $\omega=0$,
producing a weak secondary maximum at $\omega<E_F$. Thus, a weak
pseudogap is formed.


\begin{figure}
  \epsfig{figure=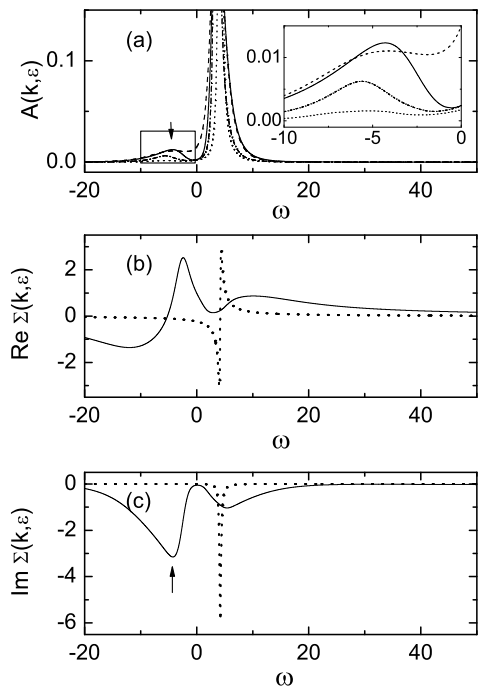,width=0.35\textwidth}
  \caption{(a) The energy dependence of the spectral function at the
    momentum indicated as point $B$ in Fig. 1. Below the Fermi energy,
    there is a small peak in the spectral function for $U=6$ and
    $T=0.1$(solid line). Dashed line: $U=6$
    and $T=0.7$, dash-dotted line: $U=4$, $T=0.1$, dotted line: $U=2$, $T=0.1$. The
    arrow points to the location of the secondary maximum. Inset:
    Zoom-in view of the solid rectangle. (b) and (c): The real and the
    imaginary part of the self-energy at $T=0.1$ for $U=6$ (solid) and
    $U=2$ (dotted). The arrow indicates the position of the extrema in
    $\Im{\Sigma(\mathbf{k},\omega)}$.}
\end{figure}

This weak pseudogap behavior is a consequence of strong Coulomb
repulsion. To show this, we also present the results with a smaller
$U=2$ at $T=0.1, 0.3, 0.5, 0.7$ in the insert of Fig. 4. Different
from the $U=6$ case, the density of states is not suppressed with
the decreasing of temperature at all(in fact there is no appreciable
difference between them, so only one line can be seen from the
figure). For a more detailed discussion, we also present the
spectral function for $U=2$ (dotted), $U=4$ (dash dotted) and $U=6$
(solid) in Fig. 5 (a), where the secondary maximum shows up
gradually with the increase of the Coulomb correlations. Note that
since the secondary maximum disappears for $U=2$, no pseudogap is
expected for this case and what for $U<2$. We refer to the
depression of the spectral weight to be a weak pseudogap here is
based on the observation that the spectral function consists of a
peak and a weak secondary maximum (not a peak). In fact, the real
part of the denominator of the Green's function has only one pole,
which can be seen from the real part of the self-energy shown in
Fig. 5 (b). It is the extremum of the imaginary part of the self
energy around $\omega=-4$ that produces the weak maximum of the
spectral function, as shown in Fig. 5 (c). In the case of a smaller
Coulomb repulsion $U$, such as $U=2$, a usual Fermi-liquid form of
the self-energy is preserved (The dotted lines in Fig. 5 (b) and
(c)), so it gives a well defined quasiparticle peak in the spectral
function (The dotted line in Fig. 5 (a)). This secondary maximum
might suggest a "shadow band",~\cite{Deisz_96} which occurs when a
short-range spin correlation is developed with the increase of
Coulomb repulsion. Therefore, the weak pseudogap behavior reported
here may be due to the spin fluctuation in the strong correlated
regime.~\cite{larger_U}

A similar weak pseudogap behavior was reported by Yada {\it et
  al.}~\cite{Yada_05} based on a multi-orbital model with the absence
of small hole pockets. Our results based on the single band model
provide further support for the sinking down of the small hole
pockets. In our opinion, the agreement between the two models suggests
that the multi-orbital effect may play a minor role in the mechanism
of the pseudogap formation.


In summary, we have studied the quasiparticle band renormalization
and the pesudogap behavior in Na$_{0.33}$CoO$_2$ based on the single
band Hubbard model in a two-dimensional triangle lattice. The
renormalization of the band structure and its anisotropy of the
$\Gamma$-$K$ and $\Gamma$-$M$ direction have been elaborated. The
estimated effective mass is consistent with the ARPES measurements.
In the meantime, a weak pseudogap behavior is reproduced, which is
explained as the result of the strong spin fluctuations. Our results
are qualitatively consistent with experiments as well as the
theoretical calculations based on a multi-orbital model.

We thank Q.-H. Wang for helpful discussions. The work was supported
by the NSFC (10525415,10474032,10429401), the RGC grants of Hong
Kong (HKU7045/04P, HKU-3/05C), and the State Key Program for Basic
Research of China (2006CB921800, 2006CB601002).

\bibliography{ref}

\begin{thebibliography}{30}
\bibitem[{lar()}]{larger_U}
\bibinfo{note}{We note that with stronger Coulomb repulsion, e.g., $U=10$, no
  qualitative difference is found. However, as expected, the
  secondary-maximum-structure in $A(\mathbf{k},\omega)$ becomes more obvious as
  well as the pseudogap effect.}
\end{thebibliography}

\end{document}